\documentclass{iopart}
\usepackage{graphicx}
\usepackage{epsf,amsfonts,amssymb,amsbsy}
\begin{document}
\title[Vector field as a quintessence partner]
{Vector field as a quintessence partner}
\author{V.V.Kiselev*\dag}
\address{*
\ Russian State Research Center ``Institute for
High Energy
Physics'', 
Pobeda 1, Protvino, Moscow Region, 142281, Russia}
\address{\dag\ 
Moscow
Institute of Physics and Technology, Institutskii per. 9,
Dolgoprudnyi Moscow Region, 141700, Russia}
\ead{kiselev@th1.ihep.su}
\begin{abstract}
We derive generic equations for a vector field driving the
evolution of flat homogeneous isotropic universe and give a
comparison with a scalar filed dynamics in the cosmology. Two
exact solutions are shown as examples, which can serve to describe
an inflation and a slow falling down of dynamical ``cosmological
constant'' like it is given by the scalar quintessence. An
attractive feature of vector field description is a generation of
``induced mass'' proportional to a Hubble constant, which results
in a dynamical suppression of 
actual cosmological constant during the evolution.
\end{abstract}


\section{Introduction}
A scalar field constitutes a basic ingredient in a preferable
theoretical treatment of two phenomena recently established
experimentally by astronomical observations \cite{SNe,WMAP}: an
inflation expansion of the Universe and a dark energy
pre-dominance at the current time.

The inflation \cite{lectures} driven by a slowly rolling scalar
field gives correct initial conditions for inhomogeneous
perturbations of a matter density, which cause the observed
anisotropy in the cosmic microwave background \cite{WMAP}. This
stage of scalar field contribution takes place at early times
before a Big Bang, so that a Harrison--Zel'dovich spectrum of
density inhomogeneity is formed.

At present, the accelerated expansion of Universe \cite{SNe} is
caused by a matter with a negative pressure, a quintessence
\cite{Q,state,trod}, which is modelled by a scalar field with an
appropriate potential, so that the visible ``cosmological term''
acquires a dynamics, that can explain an unnatural value of ``dark
energy'' scale. A scalar phantoms with a negative kinetic energy
could effectively be involved in such the studies, too
\cite{BigRip}.

There are some attempts to ascribe a dark matter in halos of
galaxies \cite{lect,univers,rev} to a scalar field, too, though
properties of such the fields at the galaxy scales should be
rather different from those of quintessence at the cosmic scale
\cite{expl,qdm,me}.

Therefore, scalar fields essentially contribute to the modern
understanding of cosmology and take control of its theoretical
progress. The only problem is an arbitrariness in a choice of
scalar potentials, which can be somehow motivated but not
certainly derived in an explicit form.

In the present paper we consider the expansion of homogeneous
isotropic universe in the presence of a vector field, which could
be a partner of quintessence. In contrast to both ref.
\cite{bertl}, wherein a gauge vector field with a global symmetry
was investigated, and ref. \cite{berg}, where a non-linear
electrodynamics led to an acceleration of universe, we study a
simple Lorentz-invariant form of lagrangian for a vector field
interacting gravitationally only,
\begin{equation}\label{v-lagr}
    {\cal L}_\textsl{v} =\xi\,\frac{1}{2}\,g^{\mu\nu}(\nabla_\mu\phi^m)\,
    (\nabla_\nu\phi^n)\,g_{mn}-V(\phi^2),
\end{equation}
where $\xi$ is a vector field signature\footnote{The signature of
metric is assigned to $(+,-,-,-)$.}, that can be normal ($\xi=-1$)
or phantomic ($\xi=+1$), respectively\footnote{Note that the
four-vector is generically composed by spin-1 and spin-0
components (see ref. \cite{ahu} for discussion on a covariant
object decomposition into components with definite spins). So,
such the companion component can cause a problem with unitarity,
which we do not consider here. However, we would implicitly
suggest, that the modes, which could non-gravitationally interact
with a matter, conserve the unitarity.}. A motivation for the
lagrangian is twofold. First, a scalar phantom with a negative
sign of kinetic energy recently studied in the context of
cosmology \cite{trod,johri} could be replaced with a
time-component of vector field in the normal mode of the
signature. Second, any gauge vector field, for example, the
abelian field ${\cal A}_\mu$ has a purely gauge component ${\cal
A}_\mu^G=\partial_\mu \omega(x)$, which does not interact with a
matter, at all. So, in the field theory the gauge invariance
preserves that the purely gauge component has a bare free
propagator invisible for the matter detectors. Moreover, this
propagator does provide the negative sign of kinetic energy.
However, the propagator could become gauge-dependent.
Nevertheless, it induces a gravitational force due to a
contribution to the energy-momentum tensor. The influence of this
force by the phantom component of vector field in a curved
space-time is under question. Thus, two mentioned items are
addressed in the present paper.

Fixing the Friedmann--Robertson--Walker metric
\begin{equation}\label{FRW}
    {\rm d}s^2 ={\rm d}t^2-a^2(t)\,[{\rm d}r^2+r^2{\rm
    d}\theta^2+r^2\sin^2\theta\,{\rm d}\phi^2],
\end{equation}
we develop a variational technique simplifying a getting the
motion equations in section 2 and then derive the evolution
equations for the scale factor $a(t)$ and the vector field, which
acquires an ``induced mass'' term with a characteristic scale of
Hubble constant\footnote{Similar effect for interacting scalar
fields was considered in \cite{Lyth}.}. In section 3 two exact
solutions are shown. The first solution gives a constant field due
to a compensation of potential variation by the induced mass term,
that can be possible for a specific potential only, so that de
Sitter space-time regime takes place. The second solution is a
free (zero potential) vector field evolving in the presence of an
actual vacuum energy, i.e. the cosmological term. However, the
contribution of this cosmological constant to the Hubble constant
gets a suppression during the evolution, so that any value can be
suppressed down to the observed scale. A slow-roll approximation
and an interaction of vector field with a scalar quintessence are
also described. The results are summarized in Conclusion.

\section{Generic equations}
In this section we develop a variational method to derive the
evolution equations in the case of FRW metric of (\ref{FRW}). So,
we give explicit expressions for the Christoffel symbols and
curvature in terms of evolving scale factor $a(t)$. A usual
variation of action over $a(t)$ results in the motion equations.
However, we show that a complete set of equations can be obtained,
if we add the invariance under the scale transformation of time.
The reason for such the procedure is quite transparent, since the
fixing of FRW metric excludes the variation of time-time component
of the metric, that can be recovered by the variation of time
scale. We check this approach by getting well-known equations for
the scalar field in the FRW background. Further, the same
technique is applied to derive the evolution equations in the case
of vector field, that is significantly simpler than a
straightforward calculation of double covariant derivatives.
\subsection{Metric values}
The interval of (\ref{FRW}) corresponds to the metric
\begin{equation}\label{metr}
    g_{tt}=1,\qquad g_{ij} =-a^2(t)\,\gamma_{ij},\qquad
    i,j=r,\theta,\phi,
\end{equation}
with the following non-zero diagonal elements:
\begin{equation}\label{gamma}
    \gamma_{rr}=1,\qquad \gamma_{\theta\theta} =r^2,\qquad
     \gamma_{\phi\phi} =r^2\sin^2\theta.
\end{equation}
The Christoffel symbols calculated by a general definition
\begin{equation}\label{Christ-def}
    \Gamma_{\mu\nu}^\lambda =\frac{1}{2}\,g^{\lambda\rho}\,
    (\partial_\mu g_{\nu\rho}+\partial_\nu g_{\mu\rho}-
    \partial_\rho g_{\mu\nu} )
\end{equation}
are given by\footnote{We use an ordinary notation for the
time-derivative by the dot-over symbol $\partial_t f(t) =\dot
f(t)$.}
\begin{equation}\label{Christof}
\begin{array}{llllllll}
  \Gamma^t_{ij} = \dot a\,a\,\gamma_{ij},&  & \displaystyle
  \Gamma^i_{tj} = \frac{\dot a}{a}\,\delta^i_j, &  & \displaystyle
  \Gamma^\theta_{r\theta} = \Gamma^\phi_{r\phi}
  =\frac{1}{r},\\[4mm]
  \Gamma^r_{\theta\theta}=-r, &&
  \Gamma^r_{\phi\phi}=-r\sin^2\theta,&&
  \Gamma^\theta_{\phi\phi}= - \sin\theta\,\cos\theta,\\[4mm]
  \displaystyle\Gamma^\phi_{\theta\phi}
  =\frac{\cos\theta}{\sin\theta},&&&&
\end{array}
\end{equation}
while the symbols are symmetric over the contra-variant indices,
and other symbols not listed above are equal to zero. In
eqs.(\ref{Christof}) we do not explicitly show the dependence of
scale factor on the time.

Then, the non-zero elements of Ricci tensor are the followings:
\begin{equation}\label{Ricci}
\begin{array}{lll}
 \displaystyle  R_{tt} = -3 \frac{\ddot a}{a},&  & \displaystyle
  R_{ij} = (\ddot a\,a+2\dot a^2)\,\gamma_{ij},
\end{array}
\end{equation}
while the scalar curvature is equal to
\begin{equation}\label{curvature}
    R = -6\left[\frac{\ddot a}{a}+\left(\frac{\dot a}{a}\right)^2
    \right].
\end{equation}
Therefor, the Einstein--Hilbert action of gravity
\begin{equation*}
S_{\textsc{g}}=-\frac{1}{16\pi G}\int R\,\sqrt{-g}\;{\rm
d}^4x=\frac{3}{8\pi G}\int(\ddot a\,a^2+\dot a^2\,a)\,{\rm d}^4x
\end{equation*}
after the integration by parts for the first term, takes the form
\begin{equation}\label{gravi}
    S_{\textsc{g}}= -\frac{3}{8\pi G}\int \dot a^2 a\;{\rm d}^4x,
\end{equation}
while the surface terms are not relevant to the variational
equations at fixed values of dynamical variables at the surface.

\subsection{Scalar field}
In the FRW metric, the action of scalar field $\phi(t)$ depending
on the time, only, is
\begin{equation}\label{scalar}
    S_{\textsl{s}} =\int a^3\,\left(\frac{1}{2}\,\dot
    \phi^2-V_{\textsl{s}}(\phi)\right)\;{\rm d}^4 x.
\end{equation}
The Euler--Lagrange equation of motion
\begin{equation*}
\frac{\partial {\mathfrak L}}{\partial
\phi}-\partial_\mu\,\frac{\partial {\mathfrak
L}}{\partial\,(\partial_\mu \phi)} =0
\end{equation*}
with the density of lagrangian ${\mathfrak L}=\sqrt{-g}\, {\cal
L}$, straightforwardly gives
\begin{equation}\label{scalar-motion}
    \ddot \phi+3\,H\,\dot \phi+\frac{\partial
    V_{\textsl{s}}}{\partial \phi} =0,
\end{equation}
where we define the Hubble constant $H=\dot a/a$.

Analogously, the Lagrange equation obtained by the variation over
the scale factor $a(t)$ including the gravity and scalar matter
actions, results in
\begin{equation*}
    -\frac{3}{8\pi G}\left[ \dot a^2-\frac{\rm d}{{\rm d}t}(2\dot
    a\,a)\right]+ 3\, a^2\,\left(\frac{1}{2}\,\dot
    \phi^2-V_{\textsl{s}}(\phi)\right)=0,
\end{equation*}
Hence, we arrive at the following equation determining the second
derivative of scale factor $a(t)$:
\begin{equation}\label{secod-a}
    2\frac{\ddot a}{a}=-H^2-8\pi G\,\left(\frac{1}{2}\,\dot
    \phi^2-V_{\textsl{s}}(\phi)\right).
\end{equation}

A key moment is a variation of the time scale, which is defined by
the following infinitesimal transformations:
\begin{equation}\label{dilaton}
    \delta_\lambda\,{\rm d}t = -\delta\lambda\,{\rm d}t,\quad
    \delta_\lambda\,\dot a =\delta\lambda\,\dot a,\quad
    \delta_\lambda\,\dot \phi =\delta\lambda\,\dot \phi,
\end{equation}
that results in the equation
\begin{equation}\label{00}
    H^2 =\frac{8\pi G}{3}\,\left(\frac{1}{2}\,\dot
    \phi^2+V_{\textsl{s}}(\phi)\right).
\end{equation}
Then, the substitution of $H^2$ in (\ref{secod-a}) by (\ref{00})
gives a usual evolution equation of scale factor governed by the
scalar field
\begin{equation}\label{second-a2}
    \frac{\ddot a}{a}=-\frac{8\pi G}{3}\,\left(\dot
    \phi^2-V_{\textsl{s}}(\phi)\right).
\end{equation}
Let us compare the above method with the standard procedure giving
the field equations of gravity, i.e. the equations
\begin{equation}
  R_{\mu\nu}-\frac{1}{2}\,R\,g_{\mu\nu} = 8\pi G\,T_{\mu\nu},
\end{equation}
where the energy-momentum tensor of scalar field is given by
\begin{equation*}
  T_{\mu\nu} = (\partial_\mu\phi)\,(\partial_\nu\phi)-{\cal
  L}\,g_{\mu\nu},
\end{equation*}
so that the time components give the energy density of the scalar
field
\begin{equation*}
    \rho_\textsl{s}=\frac{1}{2}\,\dot
    \phi^2+V_{\textsl{s}}(\phi),
\end{equation*}
while the remaining ones determine its pressure
\begin{equation*}
    p_\textsl{s} =\frac{1}{2}\,\dot
    \phi^2-V_{\textsl{s}}(\phi).
\end{equation*}
Therefore, we get ordinary evolution equations of the form
\begin{equation}\label{order}
    H^2 =\frac{8\pi G}{3}\,\rho,\qquad \frac{\ddot
    a}{a}=-\frac{4\pi G}{3}\,(\rho+3p),
\end{equation}
whereas the first equation is the positive energy condition
obtained by the variation of time component of the metric. Let us
stress that the same equation we have got under the invariance
over the time dilation.

Finally, note, that (\ref{second-a2}) is a consequence of two
other equations: (\ref{scalar-motion}) and (\ref{00}).

\subsection{Vector field}
Let us try the same method to get the equations for the gravity
and vector field.

The covariant derivative of the vector field $\phi^m$
\begin{equation}\label{covar-def}
    \nabla_\mu\phi^m = \partial_\mu\phi^m+\Gamma^m_{\mu
    n}\phi^n\equiv \phi^m_{\;\;;\,\mu}
\end{equation}
is reduced to the following non-zero components:
\begin{equation}\label{deriv}
    \phi^t_{\;;\,t}=\dot \phi_0,\quad
    \phi^i_{\;;\,j} =\delta^i_j\,\frac{\dot a}{a}\,\phi_0,
\end{equation}
where we suppose an isotropic homogeneous solution for the vector
field
\begin{equation*}
    \phi^m =\{\phi_0(t),\boldsymbol 0\}
\end{equation*}
in the FRW background. Then the action of vector field is equal to
\begin{equation}\label{action-v}
    S_\textsl{v} =\int
    a^3\left(\xi\,\frac{1}{2}\,\dot\phi_0^2+\xi\,\frac{3}{2}\,H^2\phi_0^2-V(\phi_0^2)
    \right)\,{\rm d}^4x.
\end{equation}
Therefore, the Lagrange equation for the vector field ($\delta
S_\textsl{v}/\delta \phi_0=0$) reads off
\begin{equation}\label{eq-v}
    \ddot\phi_0+3\,H\dot\phi_0-3\,H^2\phi_0+\xi\,\frac{\partial
    V}{\partial\phi_0}=0.
\end{equation}
As was for the scalar field, the Hubble constant generates a
friction $\dot \phi_0$-term in the equations of motions. In
addition, the vector field interacting with the FRW metric
possesses a feature caused by the induced mass term in
(\ref{eq-v}),
\begin{equation*}
    m^2_{\rm ind} =-3\,H^2,
\end{equation*}
which has a negative sign and depends on the dynamical value of
$H$. In the normal mode, $\xi=-1$, for the spatial vector-field
$\boldsymbol \phi$, the induced potential determines a positive
energy of $\phi_0$, while the kinetic term of time-component
$\phi_0$ has a phantom sign. Nevertheless, we continue the
consideration for both signatures.

Further, the variation over the scale factor gives
\begin{eqnarray}\nonumber
    2\,\frac{\ddot a}{a}\,(1-4\pi G\,\xi\,\phi_0^2)+H^2(1-4\pi
    G\,\xi\,\phi_0^2)&&\\ \hspace*{2.85cm}+\, 8\pi G\left(\xi\frac{1}{2}\dot\phi_0^2-V-2\xi
    H\,\phi_0\dot\phi_0\right)=0,&&
    \label{gravi-v}
\end{eqnarray}
which is more complex than that of the scalar field.

Finally, extending the time-dilation of (\ref{dilaton}) by an
additional trivial condition for the vector field
\begin{equation}\label{dilaton-v}
    \delta_\lambda\phi_0 =0,
\end{equation}
implying that the metric is not varied, we derive the third
equation (the positive energy condition),
\begin{equation}\label{eq-h^2}
    H^2 = \frac{8\pi G}{3}\,\left(\xi\,\frac{1}{2}\,\dot\phi_0^2+V+
    \xi\,\frac{3}{2}\,H^2\phi_0^2\right)\qquad\Leftrightarrow
\end{equation}
\begin{equation}\label{eq-hh}
    H^2 =\frac{8\pi
    G}{3}\,\frac{1}{1-4\pi
    G\,\xi\,\phi_0^2}\,\left(\xi\,\frac{1}{2}\,\dot\phi_0^2+V\right).
\end{equation}
Again, the form of (\ref{eq-h^2}) contains the ``mass term''
induced by the Hubble constant. Its contribution can be rewritten
in (\ref{eq-hh}), wherein we can observe an effective
gravitational constant depending on the dynamical field,
\begin{equation}\label{eff-G}
    G_{\rm eff} =G\,\frac{1}{1-4\pi G\,\xi\,\phi_0^2}.
\end{equation}

Three equations of (\ref{eq-v}), (\ref{gravi-v}) and (\ref{eq-hh})
are not independent. Indeed, (\ref{gravi-v}) is a straightforward
consequence of (\ref{eq-v}) and (\ref{eq-hh}), which gives a good
check of validity for the procedure used.

Explicitly, we rewrite (\ref{gravi-v}) in a more spectacular form,
\begin{equation}\label{gravi-v-spect}
    \frac{\ddot a}{a} =\frac{8\pi G}{3}\,\frac{1}{1-4\pi
    G\,\xi\,\phi_0^2}\, (-\xi\,\dot\phi_0^2+V+3\xi
    H\,\phi_0\dot\phi_0),
\end{equation}
which is analogous to the case of scalar field in (\ref{00}) and
(\ref{second-a2}), if only one substitutes the effective
gravitational constant of (\ref{eff-G}) and introduces an induced
pressure
\begin{equation}\label{press}
    \Delta p_{\rm ind} =-2\,\xi H\,\phi_0\dot\phi_0,
\end{equation}
which is positive, if the signature is normal $\xi=-1$, the
universe is expanded $H>0$, and the field squared is growing up,
$\partial_t(\phi_0^2)>0$. Then, common equations of (\ref{order})
with the effective gravitational constant are reproduced.

Finally, we give general equations for the evolution of flat
homogeneous isotropic universe in the presence of vector field and
a matter with the energy density $\rho_\textsl{m}$ and pressure
$p_\textsl{m}$,
\begin{equation}\label{general-1}
    H^2 =\frac{8\pi G_{\rm eff}}{3}\,\rho,
\end{equation}
\begin{equation}\label{general-2}
    \frac{\ddot a}{a}=-\frac{4\pi G_{\rm eff}}{3}\,(\rho+3 p),
\end{equation}
where
\begin{equation}\label{density}
    \rho=\rho_\textsl{m}+\xi\,\frac{1}{2}\,\dot\phi_0^2+V,
\end{equation}
\begin{equation}\label{pressure}
    p=p_\textsl{m}+\xi\,\frac{1}{2}\,\dot\phi_0^2-V-2\,\xi
    H\,\phi_0\dot\phi_0.
\end{equation}
A complete set of equations includes (\ref{eq-v}), of course.
Then, one can easily check that the conservation law for the
matter is valid as a consequence of (\ref{eq-v}),
(\ref{general-1})--(\ref{pressure}):
\begin{equation}\label{covserve}
    \dot \rho_\textsl{m}+3H(\rho_\textsl{m}+p_\textsl{m})=0.
\end{equation}

Thus, the induced mass term of the vector field in the FRW metric
provides a rich phenomenology, two examples of which are shown in
the next section.

\section{Some 
solutions}
The consideration of evolution is essentially simplified, if we
neglect the external matter.

\subsection{Constant field}
First, let us describe the case of constant field, i.e. we put
\begin{equation*}
    \dot \phi_0\equiv 0\quad\Rightarrow\quad \dot G_{\rm eff} =0.
\end{equation*}
Then we get
\begin{equation}\label{e1}
    3\,H^2\phi_0=\xi\,\frac{\partial
    V}{\partial\phi_0},
\end{equation}
and
\begin{equation}\label{e1'}
    H^2 =\frac{8\pi
    G}{3}\,\frac{1}{1-4\pi
    G\,\xi\,\phi_0^2}\,V,
\end{equation}
so that the potential should have a form
\begin{equation}\label{e1"}
    V=V_0\,\frac{1}{1-4\pi G\,\xi\,\phi_0^2}.
\end{equation}
Then, the Hubble constant is independent of time,
\begin{equation*}
    \dot H \equiv 0,
\end{equation*}
and its square is positive, if $V_0>0$,
\begin{equation*}
    H^2=\frac{8\pi G^2_{\rm eff}}{3 G}\,V_0.
\end{equation*}
At $\xi=-1$ we avoid a singularity in the potential. In this case
an actual value of primary cosmological constant can be suppressed
as
\begin{equation*}
    V_0^{\rm eff} =V_0\,\frac{1}{(1+4\pi G\,\phi_0^2)^2},
\end{equation*}
if the field value is large in comparison with the Planck mass,
$m^2_{\rm Pl}=1/4\pi G$.

\subsection{Cosmological constant}

Second, we make a field potential to be a trivial constant
\begin{equation*}
    V\equiv V_0.
\end{equation*}
Then, the filed could asymptotically evolve with a zero
acceleration
\begin{equation*}
    \ddot \phi_0\equiv 0,\qquad\mbox{at } \phi_0^2\gg m^2_{\rm
    Pl}.
\end{equation*}
Indeed, we have got
\begin{equation*}
    \dot \phi_0=H\,\phi_0\quad\Rightarrow
\end{equation*}
\begin{equation*}
    H^2\left(1-\xi\,\frac{\phi_0^2}{m^2_{\rm Pl}}\right)
    =\frac{2}{3m^2_{\rm
    Pl}}\,\left(V_0+\xi\,\frac{1}{2}\,H^2\phi_0^2\right)\quad
    \Rightarrow
\end{equation*}
\begin{equation*}
    H^2\approx-\xi\,\frac{V_0}{2\phi_0^2}.
\end{equation*}
Moreover, the acceleration parameter is
\begin{equation*}
    \frac{\ddot a}{a}\,\frac{1}{H^2} \equiv
    q=\frac{1}{1-\xi\,\phi_0^2/m^2_{\rm Pl}}\approx 0.
\end{equation*}
The asymptotic solution at $\phi_0^2\gg m^2_{\rm Pl}$ reads off
\begin{equation}\label{asymp}
    \phi_0(t) = (t-t_*)\,\sqrt{-\xi\,\frac{V_0}{2}},
\end{equation}
where $t_*$ is an integration constant. Then, we find
\begin{equation}\label{h-t}
    H\,(t-t_*) =1,
\end{equation}
and the acceleration is equal to zero.

No doubt, the above two solutions can be considered as
preliminaries for a phenomenological treatment, when one should
include the matter and radiation as well as a scalar quintessence
or inflation field. Nevertheless, we see that the vector field
could serve as an original source for both regimes: the inflation
and current acceleration in the universe expansion.

\subsection{Slow-roll approximation}
Consider the case, when one can approximately neglect higher
derivative terms for the vector field, i.e. the kinetic energy in
comparison to the potential one and a variation of the kinetic
energy in the field equations. So, we get the following slow-roll
conditions:
\begin{equation*}
    \epsilon_\textsl{v} =\left|\frac{\dot\phi_0^2}{2 V}\right|\ll
    1, \qquad
    \eta_\textsl{v} =\left| \frac{\ddot\phi_0}{3
    H\dot\phi_0}\right|\ll 1.
\end{equation*}
Then, the field equations are reduced to
\begin{equation*}
    H^2=\frac{2}{3m^2_{\rm Pl}}\,\frac{V}{1-\xi\,\phi_0^2/m^2_{\rm
    Pl}},\qquad
    3H\dot \phi_0-3H^2\phi_0+\xi\,V^\prime=0,
\end{equation*}
where $V^\prime=\partial V/\partial\phi_0$. Then
\begin{equation*}
    \epsilon_\textsl{v} =\frac{1}{12}\,|m^2_{\rm Pl}-\xi\,\phi_0^2|
    \left(\frac{V^\prime}{V}-2\xi\,\frac{\phi_0}{m^2_{\rm Pl}-\xi\,\phi_0^2}\right)^2,
\end{equation*}
\begin{equation*}
    \eta_\textsl{v} =\left|\frac{1}{12}\,(m^2_{\rm
    Pl}-\xi\,\phi_0^2)\,
    \left(\frac{V^\prime}{V}+2\xi\,\frac{\phi_0}{m^2_{\rm Pl}-\xi\,\phi_0^2}\right)^2\right.
\end{equation*}
\begin{equation*}\hspace*{3.3cm}\left.
    +\,
    \frac{\xi}{3}-\frac{1}{6}\,\frac{V^{\prime\prime}}{V}\,(m^2_{\rm
    Pl}-\xi\,\phi_0^2)\right|.
\end{equation*}
So, in the limit of cosmological constant at $\xi=-1$ we get
\begin{equation*}
    \epsilon_\textsl{v}=\frac{1}{3}\,\frac{\phi_0^2}{m^2_{\rm Pl}+\phi_0^2} \leqslant
    \frac{1}{3},\qquad
    \eta_\textsl{v} =\frac{1}{3}\,\frac{m^2_{\rm Pl}}{m^2_{\rm Pl}+\phi_0^2} \leqslant
    \frac{1}{3},
\end{equation*}
therefore, the slow-roll approximation holds good. Furthermore, in
this case
\begin{equation*}
    \frac{\dot \phi_0}{\phi_0} = H \quad \Rightarrow\quad
    \phi_0(t)=\phi_*\,a(t),
\end{equation*}
and the equation
\begin{equation*}
    H^2=\frac{2 V_0}{3m^2_{\rm
    Pl}}\,\frac{1}{1+a^2\frac{\phi^2_*}{m^2_{\rm Pl}}}
\end{equation*}
can be exactly integrated out, so that at small $a$ we get the
exponential inflation with a constant Hubble rate, $a\sim
\exp(t\sqrt{2V_0/3m^2_{\rm Pl}})$, while at large $a$ the
asymptotic behavior given in the previous subsection takes place.
Thus, the vector field can provide a natural unification of the
inflation and ordinary expansion, even at the simplest constant
potential.

\subsection{Matter and vector field}
The conservation law for the matter with a constant parameter of
state $w$
\begin{equation*}
    p_\textsl{m}=w\rho_\textsl{m}
\end{equation*}
results in
\begin{equation*}
    \rho_\textsl{m}=\frac{\rho_*}{a^n},\quad n=3(1+w).
\end{equation*}
If the vector field has a trivial potential, i.e. a constant, then
it again evolves as
\begin{equation*}
    \phi_0 \sim a,
\end{equation*}
so that at small $a$, when the matter dominates, we get a standard
cosmology, while at large $a$ we find a dynamical suppression of
cosmological constant.

\subsection{Interacting scalar and vector fields}
It is an easy task to derive the evolution equations for the case
of vector field interacting with a scalar field $\phi$. We
restrict ourselves by the consideration of slow-roll approximation
in the case, when the interaction is given by a potential, only,
without any dynamical terms, i.e. the derivatives of the fields.
Moreover, we take the potential of the form
\begin{equation*}
    V(\phi_0,\phi) = U(\phi)(1+4\pi G\,\phi_0^2).
\end{equation*}
Then the slow rolling gives\begin{equation}\label{slow-h}
    H^2=\frac{8\pi G}{3}\,U,
\end{equation}
\begin{equation}\label{slow-p}
    3H\,\dot\phi_0-3H^2\phi_0-8\pi G\,\phi_0=0,
\end{equation}
\begin{equation}\label{slow-s}
    3H\,\dot\phi+\frac{\partial U}{\partial \phi}\,(1+4\pi
    G\,\phi_0^2)=0,
\end{equation}
where we put the signature for the vector field $\xi=-1$. The
substitution of (\ref{slow-h}) into (\ref{slow-p}) leads to
\begin{equation*}
    \dot\phi_0 = 2H\,\phi_0\quad \Rightarrow \quad \phi_0(t)
    =\phi_*\,a^2(t).
\end{equation*}
So, the vector field evolves quadratically with the scale factor.
In the regime, when the vector field is small in comparison to the
Planck mass, $4\pi G\,\phi_0^2\ll 1$, the running of scalar
quintessence and scale factor evidently repeats the situation with
no any vector field. Therefore, we further consider the regime of
\begin{equation*}
    4\pi G\,\phi_0^2\gg 1,
\end{equation*}
and specify the quintessence potential as a falling homogeneous
function
\begin{equation}\label{homo}
    U=U_0\,\frac{M^{2n}}{\phi^{2n}},
\end{equation}
where $M$ is a scale of mass, and $n$ is a positive number.
Further,
\begin{equation*}
    \dot\phi = H\,\phi^\prime, \qquad
    \phi^\prime=\frac{\partial\phi}{\partial\ln a}.
\end{equation*}
Then in the large vector field regime we get
\begin{equation}\label{ss}
    2\,\phi^\prime =-\frac{\partial\ln U}{\partial
    \phi}\,\phi_*^2\,a^4,
\end{equation}
so that for the homogeneous potential we obtain
\begin{equation*}
    2\phi\phi^\prime =2n\phi_*^2\,a^4,
\end{equation*}
that gives
\begin{equation*}
    \phi^2 =\frac{n}{2}\,\phi_*^2\,a^4+\phi_c^2,
\end{equation*}
where $\phi_c$ and $\phi_*$ are constants of integration. If
$\phi_c$ dominates we have approximately got a constant scalar
field and a constant Hubble rate of inflation. Otherwise, at $a\to
\infty$ both fields proportionally evolve
\begin{equation*}
    \phi=\phi_0\,\sqrt{n/2},
\end{equation*}
and
\begin{equation*}
    H^2=U_0\,\frac{8\pi
    G}{3}\,\frac{M^{2n}}{\phi_*^{2n}}\,\frac{1}{a^{4n}},
\end{equation*}
that can be easily integrated out,
\begin{equation*}
    a\sim t^{{1}/{2n}},\qquad H\sim \frac{1}{t}.
\end{equation*}
These relations can be compared with the standard quintessence
evolution obtained by $\phi_0=0$, so that
\begin{equation*}
    \ln a \sim t^{2/(2+n)},\qquad H \sim t^{-n/(2+n)}.
\end{equation*}
Such the partnership of vector field with the scalar quintessence
could be involved in the dynamics of universe today and in future,
since a current cosmological constant
\begin{equation*}
    \Lambda= U_0\,\frac{M^{2n}}{\phi_c^{2n}}
\end{equation*}
can be naturally small even at $U_0\sim M^4$, $M\sim
M_{\rm\uppercase{gut}}$, if $\phi_c\gg M_{\rm\uppercase{gut}}$ and
$n\gg 1$, or at $M$ giving the supersymmetry breaking scale. In
addition, this ``cosmological constant'' will dynamically falling
down in future.

\section{Conclusion}
In this paper we have derived generic equations for the evolution
of flat homogeneous isotropic universe driven by a vector field.

We have found that the expansion induces a dynamical mass term for
the vector field. Such the term can be treated as a source of
running gravitational constant. This characteristic property can
lead to a dynamical suppression of primary cosmological constant.
The regimes of de Sitter expansion as well as a slowly rolling
acceleration have been demonstrated. So, the vector field could
serve not only as a quintessence partner, but it could compete
with the scalar quintessence as an origin of dark energy in the
universe.

Motivations for the consideration of vector field dynamics in the
theory of gravitation could be manyfold. So, as we have already
mentioned, a success of scalar quintessence is significantly based
on a freedom in a choice of scalar potentials, which can be
motivated in some underlying theories. Nevertheless, there are
several aspects, which seem to point to a way for a possible
compelling extensions. First, a phantomic quintessence modelled by
a scalar field with a negative kinetic term can be {\it
automatically} included in the consideration by introduction of
non-gauge vector field as given by the lagrangian of
((\ref{v-lagr}). Second, a present day quintessence has a
characteristic scale of mass about the Hubble rate, while such the
induced mass term is {\it automatically} generated by covariant
derivatives for the non-gauge vector field. Third, scalar fields
are involved in the explanation of flat rotation curves in dark
galactic halos, so that the fields compose a triplet of
monopole-like configuration \cite{expl} i.e. the spherically
symmetric static field is proportional to the radius-vector:
$\phi^a\propto \boldsymbol x$. This fact implies that explicit
introduction of corresponding vector field could be promising in
the aspect of dark matter in galaxies\footnote{This issue is
considered in \cite{mee}.}. Next, if the vector field state at
zero point $\phi^m\equiv 0$ is unstable, then the vector field
could make the time arrow in the universe expansion.

We have considered several toy models, which have confirmed the
physical effects expected due to the introduction of vector field
as listed above. However, there are strong phenomenological
constraints, which pose restrictions on a possible model. Indeed,
the most strong constraints follow from the variation of Newton's
constant and anomalous gravitational production of light fields
during an inflation.

As for the variation of Newton's constant, we note that the vector
field can cause a time-dependence, which is different from the
variation of constant universality. So, the time-dependence is
experimentally suppressed \cite{Will} at the limit of
    \begin{equation*}
    \frac{|\dot G|}{G} \lesssim \mbox{few}\times
    10^{-11}\;\mbox{year}^{-1},
\end{equation*}
which is in agreement with a constraint following from the
nucleosynthesis: the variation of Newton's constant at the time of
nucleosynthesis in comparison with the present day value should be
less than
\begin{equation*}
    \frac{\Delta G}{G}\lesssim 30\%,
\end{equation*}
so that dividing by the Universe age about $13.7$ billion years,
we get a similar estimate. Therefore, the corresponding evolution
of temporal component $\phi_0$ should be essentially suppressed.
For instance, a preferable choice is a model with a vector field
settled in a stable point, so that the evolution can be neglected,
if the field mass is much greater than the Hubble rate.

An analogous note has to be made on the vector field mass much
greater than the Hubble rate, because of an anomalous
gravitational production of fields with a similar mass scale as
the vector field under consideration. Indeed, Felder, Kofman and
Linde \cite{FKL} have found that a moduli field coupled with the
gravity can be copiously produced at early stages of inflation, if
an effective mass of moduli has a dependence on the Hubble rate,
\begin{equation*}
    m^2_{\rm eff} =m^2+ c^2\,H^2,
\end{equation*}
with a constant $c$ and a `bare' mass $m$. The production can be
suppressed, if $c\gg 1$, or $H\ll 10^{14}$ GeV, or $m\gg H$. Then,
preferably we should again adopt the constraint
\setcounter{equation}{43}
\begin{equation}\label{m>H}
    m\gg H
\end{equation}
during inflation, that implies that the vector field is, in
practice, at rest with a mass about the Planck scale.

Thus, in cosmology a variety of vector field potential is actually
restricted by the constraint of vector field stabilization with a
scale about the Planck mass, while other models probably are
removed since they lead to unacceptably large variation of
Newton's constant and anomalous gravitational production of light
vector fields during the inflation.

In addition, to the moment there is no any fundamental physics
framework for the introduction of extra vector fields in the
phenomenology, to my knowledge. Nevertheless, one should point to
that gauge vector fields of the Standard Model have longitudinal
components, which are decoupled from the interactions with matter
fields. The gauge invariance of interaction guarantees that the
propagator of longitudinal component is not renormalized, i.e. it
remains bare under the gauge interactions, and the longitudinal
mode is never seen by the matter. In this respect the non-gauge
lagrangian of (\ref{v-lagr}) differs from the gauge invariant one
by a specific gauge fixing term. This difference is never seen by
the gauge-interacting matter. A natural question is whether this
mode does play any role in the gravity or not (for an abelian
field under consideration). So, the present paper can be useful in
this aspect, too.

Phenomenological models are beyond the scope of this paper, so
that the toy examples considered cannot answer a question whether
the vector field could be a useful partner of quintessence in
practice or not, but we have shown that the vector field could be
the partner in principle. Advantages of vector field partnership
with the quintessence could be model-dependent.

This work is partially supported by the grant of the president of
Russian Federation for scientific schools NSc-1303.2003.2, and the
Russian Foundation for Basic Research, grant 04-02-17530.



\end{document}